\documentclass[journal,twoside,web]{ieeecolor}
\usepackage{lcsys}
\usepackage{cite}
\usepackage{amsmath,amssymb,amsfonts}
\usepackage{graphicx}
\usepackage{textcomp}
\def\BibTeX{{\rm B\kern-.05em{\sc i\kern-.025em b}\kern-.08em
    T\kern-.1667em\lower.7ex\hbox{E}\kern-.125emX}}
\newcommand{\argmin}[1]{\operatorname{arg}\underset{#1}{\operatorname{min}}\,}

\newcommand{\norm}[1]{\left\lVert#1\right\rVert}

\DeclareMathOperator{\diag}{diag}

\usepackage[mathscr]{eucal}
\usepackage{amsmath}
\usepackage{algorithm}		
\usepackage{algpseudocode} 	
\usepackage{tabularx} 	
\usepackage{makecell}  	
\usepackage{booktabs}  	
\usepackage[dvipsnames, rgb]{xcolor}	
\usepackage{tikz}
\usepackage{subcaption}
\usepackage{pgfplots}
\usepackage[labelsep=period, font=small]{caption}
\pgfplotsset{compat=1.18}

\newtheorem{definition}{Definition}
\newtheorem{lemma}{Lemma}

\begin{document}
\title{Bayesian Knowledge Transfer for a Kalman Fixed-Lag Interval Smoother}
\author{Ondřej Skalský and Jakub Dokoupil
	\thanks{The work has been performed in the project A-IQ Ready: Artificial Intelligence using Quantum measured Information for realtime distributed systems at the edge (Grant No. 101096658/9A22002) and was co-funded by grants from the Ministry of Education, Youth and Sports of the Czech Republic and the Chips Joint Undertaking.
	Furthermore, this work was supported in part by the Czech Science Foundation under Grant 23-06476S, in part by the European Union through the project Robotics and Advanced IndustProduction (Grant No. CZ.02.01.01/00/22\_008/0004590), in part by the infrastructure of RICAIP, which has received funding from the European Union’s Horizon 2020 research and innovation programme (Grant agreement No. 857306), and in part by the Ministry of Education, Youth and Sports under OP RDE Grant agreement No. CZ.02.1.01/0.0/0.0/17\_043/0010085.}
	\thanks{The authors are with the Faculty of Electrical Engineering and Communication and the Central European Institute of Technology, Brno University of Technology, 612 00 Brno, Czech Republic (e-mail: \linebreak ondrej.skalsky@ceitec.vutbr.cz; jakub.dokoupil@ceitec.vutbr.cz).}
}

\pagestyle{empty}		
\maketitle
\thispagestyle{empty}	

\begin{abstract}
	A Bayesian knowledge transfer mechanism that leverages external information to improve the performance of the Kalman fixed-lag interval smoother (FLIS) is proposed.
	Exact knowledge of the external observation model is assumed to be missing, which hinders the direct application of Bayes' rule in traditional transfer learning approaches.
	This limitation is overcome by the fully probabilistic design, conditioning the targeted task of state estimation on external information.
	To mitigate the negative impact of inaccurate external data while leveraging precise information, a latent variable is introduced.
	Favorably, in contrast to a filter, FLIS retrospectively refines past decisions up to a fixed time horizon, reducing the accumulation of estimation error and consequently improving the performance of state inference.
	Simulations indicate that the proposed algorithm better exploits precise external knowledge compared to a similar technique and achieves comparable results when the information is imprecise.
\end{abstract}

\begin{IEEEkeywords}
	Bayesian knowledge transfer, fixed-lag interval smoothing, state estimation, fully probabilistic design.
\end{IEEEkeywords}

\section{Introduction}\label{sec:introduction}
\IEEEPARstart{W}{ith} applications across many domains, such as deep \cite{Zhu:21} and reinforcement \cite{Torr:09},\cite{Zhu:23} learning, computer vision, transportation \cite{Zhu:21}, probabilistic learning, and signal processing, transfer learning has become a key direction in machine learning.
A central concern in transfer learning is to improve the learning of a target task in the target domain by transferring external knowledge from a related but different domain \cite{Zhu:21}.
The transfer mechanism design typically involves a trade-off between aggressivity and robustness. 
Aggressive approaches can lead to significant learning improvements within the target domain but comprise the risk of substantial negative transfer when external knowledge is imprecise.
In contrast, robust methods ensure that unrelated external data do not cause negative transfer, even though their ability to fully exploit precise and relevant information is usually limited \cite{Torr:09}. 
In Bayesian transfer learning in particular, if an explicit model describing the stochastic relation of external information to a target quantity of interest is available, then Bayes' rule is the consistent mechanism to process the knowledge, i.e., to refine the probability density function (pdf) of the target quantity in light of the external information. 
The experimenter's knowledge, however, rarely suffices to construct such models infallibly. 
The problem of optimally incorporating external information into the target inference process in these scenarios is undoubtedly one of the most important open questions in the empirical sciences \cite{Karn:16}.

Among the decision-making strategies, we address this difficulty by adopting the fully probabilistic design (FPD) \cite{Karn:96}---an axiomatized and formally justified \cite{Guy:12} extension of the principle of minimum cross-entropy \cite{Sho:80}. 
In the FPD, the experimenter's preferences about the model are expressed via an ideal pdf. 
Then, within the external knowledge-constrained set of admissible pdfs, 
the one minimizing Kullback-Leibler divergence (KLD) \cite{Kull:51} to the ideal model determines the posterior pdf \cite{Karn:16}.
Out of the numerous FPD applications \cite{Karn:16}, those that stand the closest to the problem of our concern---the state estimation of a linear Gaussian state-space model---were addressed in \cite{Quinn:18,Pap:18,Pap:19,Pap:21} and extended in \cite{Pap:stud} to a Student's model.
In \cite{Quinn:18,Pap:18,Pap:19,Pap:21}, the authors assumed an autonomous system, employing FPD in transfer learning between a pair of Kalman filters (KFs) \cite{Kal:60}, with the external knowledge embodied by an output predictive pdf of an external KF.
Thus, besides the target task, an additional filter had to be implemented for the external domain, requiring not only additional computational costs but also a complete probabilistic description of the external state-space model.
Moreover, \cite{Quinn:18} and \cite{Pap:18} required informal modifications to achieve robust transfer, and \cite{Pap:18} yielded a non-recursive solution similar to a smoothing structure \cite{RTS:65}.
The informal adaptations were addressed in \cite{Pap:19} by introducing a scalar latent variable, and in \cite{Pap:21} by invoking a hierarchical Bayesian transfer learning structure \cite{Karn:16}, which required a computationally extensive Monte Carlo integration. 

To the best of the authors’ knowledge, our article proposes the first application of the FPD in transfer learning for a fixed-lag interval smoother (FLIS)\cite{Moore:73}---in a sliding mode operating modification of the conventional fixed-interval smoother \cite{RTS:65}---in the state estimation of a non-autonomous Gaussian state-space system.	
Compared to filtering techniques, online smoothers generally provide more accurate estimates at the expense of some delay.
Consequently, smoothers find application in a wide range of areas, such as signal processing, target tracking, navigation, and communication \cite{Xu:21}.
Remarkably, when employing loss-based decision-making techniques, FLISs naturally tend to reduce the accumulation of estimation error compared to filters, as the decisions are retrospectively refined over a defined time horizon.
Thus, FLISs may outperform filters even in the \textit{filtering regime}---reporting only the up-to-date state estimate among all others.

The approach of weighting the transfer rate by a latent variable, as introduced in \cite{Pap:19}, has significantly influenced the direction of our research. 
Although the method in \cite{Pap:19} is robust, it lacks the ability to fully leverage precise external data.
Inspired by this idea, we also adopt a latent-variable-based weighting strategy. However, our method departs from the prior research in several key aspects.
Besides inheriting the merits of the FLIS, the proposed algorithm, unlike \cite{Quinn:18, Pap:18, Pap:19, Pap:21}, requires no external information other than output observations, and additionally obviates the need for preprocessing external data.
Owing to a conceptually different modeling approach and processing strategy, the proposed FPD-based design better exploits precise external data while retaining robustness to imprecise ones.
Moreover, for a multidimensional system output, the presented algorithm is capable of leveraging even partially corrupted external data, as the weighting variable is not a scalar but a matrix.

\textbf{Notation.}
The notation $f(\cdot)$ is reserved for a known pdf distinguished by its argument and, optionally, by a subscript; $\breve{f}(\cdot)$ represents a variational (unknown) pdf; 
$\hat{f}(\cdot)$ denotes the optimal choice of a variational pdf with respect to a given criterion; 
$x^{*}$ symbolizes the range of $x$; $\mathring{x}$ defines the number of elements in a countable set $x^{*}$; 
$x_{z|t}$ refers to a quantity $x$ expected to prevail at a discrete time $z$, given target data up to a time $t$;
and, by extension, $x_{z|^t_e}$ refers to a quantity $x$ expected to prevail at a discrete time $z$, given target data up to a time $t$ and external data up to a time $e$. 
Further, $\propto$ indicates equality up to a normalizing constant; $'$ denotes transposition; $\equiv$ means equality by definition; $\otimes$ represents the Kronecker product; 
$\circ$ stands for the Hadamard product;
$\mathrm{tr}(\cdot)$ is the trace operator; 
$|\cdot|$ symbolizes the absolute value of the determinant;
$\norm{\cdot}_2$ represents the Euclidean norm;
$I_n$ is the identity matrix of size $n\times{}n$; $O_{n,m}$ is an $n\times m$ zero matrix; $\epsilon_i^n$ denotes the $i$th column of the identity matrix $I_n$; $\overline{\epsilon}_i^n$ indicates the $i$th row of the identity matrix $I_n$; $\diag(\cdot)$ forms a diagonal matrix from a given vector; and the mathematical expectation of an arbitrary function $g(x)$ with respect to the pdf $f(x)$ is expressed as $\mathcal{E}_{f(x)}[g(x)] = \int_{x^*}{g(x)f(x)\mathrm{d}x}$. 
Lastly, 
$
\mathcal{N}(x|\hat{x},P) \propto  \exp
\bigl\{
-(x-\hat{x})'P^{-1}(x-\hat{x})/2
\bigr\}
$
denotes the normal distribution of a vector $x \in \mathbb{R}^{\mathring{x}}$ with parameters $\{\hat{x},P \}$; and
$
i\mathcal{W}(\Xi|\Sigma,\nu) \propto 
|\Xi|^{-(\nu+\mathring{y}+1)/2}
\exp
\bigl\{
-\mathrm{tr}(\Xi^{-1}\Sigma)/2
\bigr\}
$
is the inverse-Wishart distribution of a symmetric positive definite matrix $\Xi$ of dimension $\mathring{y} \times \mathring{y}$, with parameters $\{\Sigma, \nu\}$.

\section{Algorithm Design}
We suppose that a noiseless input $u_k \in \mathbb{R}^{\mathring{u}}$ and a noisy output $y_{\mathrm{T};k} \in \mathbb{R}^{\mathring{y}}$ are observed on a stochastic target MIMO state-space system at discrete time instants $k\in{}k^{*}\equiv\{1, 2,\ldots,\mathring{k}\}$.
The input-output pairs collected from time $1$ to time $k$ form the data record $\mathscr{D}_{1}^k\equiv \{u_i,y_i\}_{i=1}^k$. 
Assuming the Markov property \cite{Kol:33} and the independence of the process and measurement noise, the state-space system is specified by the observation model \eqref{eq:model_functional_y} and the time evolution model \eqref{eq:model_functional_x}
	\begin{align}
		\label{eq:model_functional_y}
		y_{\mathrm{T};k} &\sim f(y_{\mathrm{T};k}|x_k)  \equiv \mathcal{N}(y_{\mathrm{T};k}|Cx_k,R),
		\\
		\label{eq:model_functional_x}
		x_{k+1}  &\sim f(x_{k+1}|x_{k},u_{k}) \equiv \mathcal{N}({x_{k+1}}|Ax_{k}+Bu_{k},Q).
	\end{align}
Here, $x_k \in \mathbb{R}^{\mathring{x}}$ is the state variable to be estimated. 
The parameters of the linear Gaussian model~\eqref{eq:model_functional_y}--\eqref{eq:model_functional_x} are the appropriately dimensioned state transition $A$, input $B$, observation $C$, state noise $Q$, and observation noise $R$ matrices.
For notational convenience, the conditioning on $\{A, B, C, Q, R\}$ is omitted throughout the article.
We assume that \( R \) is diagonal, implying conditional independence among the elements of \( y_{\mathrm{T};k} \).	This allows us to formulate a matrix-inversion-free algorithm with reduced computational demands.

\subsection{Kalman Fixed-Lag-Interval Smoother}
The concern of the FLIS is to jointly estimate the time sequence of the states, grouped into the augmented state vector
\begin{align*}
	X_k \equiv 	\bigl[x_k', x_{k-1}',\ldots,x_{\max(k-L,1)}'\bigr]'.
\end{align*}
The parameter $L$ determines the fixed-lag smoothing delay.
Initialized at $k=1$ with the experimenter's initial knowledge provided in the form of the statistics $\{ \hat{X}_{1|0}, P_{1|0} \}$ of the prior $\mathcal{N}(X_1 | \hat{X}_{1|0}, P_{1|0})$, the recursive smoothing is accomplished by repeatedly performing the data \eqref{eq:DS} and time \eqref{eq:TS} steps, also known as the state prediction and correction, respectively:
\begin{align}
	\label{eq:DS}
	&f(X_k|\mathscr{D}_1^{k}) 
	= \mathcal{N}\big(X_k\big|\hat{X}_{k|k},P_{k|k}\big)
	\\
	&
	\quad\quad
	= f(X_k|\mathscr{D}_1^{k-1}) f(y_{\mathrm{T};k}|x_k)/f(y_{\mathrm{T};k}|\mathscr{D}_1^{k-1}),
	\notag
	\\
	\label{eq:TS}
	&f(X_{k+1}|\mathscr{D}_1^{k}) =
		\mathcal{N}\big(X_{k+1}\big|\hat{X}_{k+1|k},P_{k+1|k}\big) \\
	&
	\quad\quad
	= 
\left\{
\!\!\!
\begin{array}{ll}
	\int_{x^*} f(x_{k+1}|x_{k},u_{k}) f(X_{k}|\mathscr{D}_1^{k}) \mathrm{d}x_{k-L}
	& \quad k > L \\[0.5em]
	f(x_{k+1}|x_{k},u_{k}) f(X_{k}|\mathscr{D}_1^{k})
	& \quad k \leq L.
\end{array}
\right.
	\notag
\end{align}
	At this stage, we introduce auxiliary variables to simplify the forthcoming expressions: $w \equiv \min(k, L+1)$ denotes the smoothing window size at time $k$; $q \in q^*$ indexes the time steps within the smoothing horizon $q^* \equiv \{k-w+1,\dots,k\}$; and $l \equiv \min(k, L)$.
Let us begin by elucidating the distinction between the FLIS and the conventional fixed-lag smoother (FLS). 
The FLS's estimation result is assumed \cite{Med:67} to be the smoothed marginal pdf $f(x_{k-w+1}|\mathscr{D}_1^k)$. In contrast, the presented FLIS yields the joint pdf \( f(X_k | \mathscr{D}_1^k) \).
Thus, the FLIS inherently preserves the states' cross-correlations across the time horizon $q^*$, which is computationally as well as notationally advantageous for the presented knowledge transfer design.

\subsection{FPD-Optimal Knowledge Transfer for FLIS}
By introducing a hidden variable $\Xi$, we extend the observation model \eqref{eq:model_functional_y}, which is valid in the non-transfer scenario, to the general observation model
\begin{align}
	\label{eq:y_general}
	y_k \sim f(y_k|x_k,\Xi) \equiv \mathcal{N}(y_k|Cx_k,R\Xi).
\end{align}
The scale matrix $\Xi$ of an appropriate dimension arbitrarily adjusts \pagebreak the variance of the fictive observation $y_k \in \mathbb{R}^{\mathring{y}}$, making \eqref{eq:y_general} universal for both non-transfer and transfer scenarios.\footnote{
	Given $x_k$ and $\Xi = I_{\mathring{y}}$, the observation $y_{\mathrm{T};k}$ can be interpreted as a certain realization of $y_k$:
	$	   \int_{y^*}
	\delta(y_k - y_{\mathrm{T};k}) f(y_k|x_k,\Xi=I_{\mathring{y}})\, \mathrm{d}y_k = f(y_{\mathrm{T};k}|x_k).
	$
}
Analogously to $y_{\mathrm{T};k}$, the elements of $y_k$ are assumed to be conditionally independent, implying that $\Xi$ is diagonal.
We assume that the external knowledge is available in the form of observations forming the data record $\mathrm{E}_1^k \equiv \{ y_{\mathrm{E};i} \in \mathbb{R}^{\mathring{y}} \}_{i=1}^k$.
However, the linkage between $\mathrm{E}_1^k$ and the quantity of interest $X_k$ is absent, preventing the direct application of Bayes' rule. 	
To overcome this limitation, we define the predictive pdf
\begin{align}
	\label{eq:external_predictor}
	f_\mathrm{E}(y_k|y_{\mathrm{E};k},\Xi) \equiv \mathcal{N}(y_k|y_{\mathrm{E};k},\Xi),
\end{align}
which uses the scale matrix $\Xi$ to quantify the uncertainty in interpreting $y_{\mathrm{E};k}$ as a realization of $y_k$. 
This formulation allows applying the FPD, as proposed in \cite{Karn:17}.

\begin{definition}
	Let $B$ denote a set comprising observations and a transfer learning prior on hidden variables of interest.
	The designer's preferences about $B$ are specified by an ideal pdf, $f_\mathrm{I}(B)$.
	The admissible models $\breve{f}(B) \in \breve{f}^*(B)$ are defined by a constrained set, $\breve{f}^*(B)$, which reflects the available knowledge, model structure, or other design limitations.
	Then, the FPD-optimal pdf
	\begin{align}\label{eq:FPD}
		\hat{f}(B)
		\equiv \argmin{\breve{f}(B)}
		\mathcal{D}\bigl(\breve{f}(B)\big\| f_\mathrm{I}(B)\bigr)
	\end{align}
	is prescribed as the one that minimizes the KLD
	\begin{align}\label{eq:KLD}
		\mathcal{D}\bigl(\breve{f}(B)\big\| f_\mathrm{I}(B)\bigr)
		\equiv
		\int_{B^*}
		\!\!
		\breve{f}(B) \ln \Biggl( 
		\frac{\breve{f}(B)}{f_\mathrm{I}(B)}
		\Biggr)
		\mathrm{d}
		B
	\end{align}
	to the defined ideal model $f_\mathrm{I}(B)$. 
	For a more detailed discussion of the FPD, we refer the reader to \cite{Guy:12}.
\end{definition}

To restrict the computational complexity of the optimization problem \eqref{eq:FPD}, we limit the considered observations to the $w$ 	most recent fictive ones: $\mathrm{Y}_{k-w+1}^k \equiv \{y_q\}_{q\in q^*}$.
Furthermore, since $\Xi$ effectively quantifies the uncertainty of the external information, we relax the assumption of its exact knowledge and treat it as a random variable.
In line with these considerations, we define $B \equiv \{ \mathrm{Y}_{k-w+1}^k, X_k, \Xi \}$.

Let the admissible models of $B$ be restricted by the external predictor \eqref{eq:external_predictor}~\cite{Karn:17} and a factorization constraint:	
\begin{align}
	\label{eq:varset}
	&
	\breve{f}(B)
	\equiv
	\breve{f}(\mathrm{Y}_{k-w+1}^k,X_k,\Xi|\mathscr{D}_1^k,\mathrm{E}_1^{k})
	\\
	&
	\quad
	\equiv
	\!\!\!\!\!
	\prod_{q = k-w+1}^k 
	\!\!\!\!\!
	f_\mathrm{E}(y_q|y_{\mathrm{E};q},\Xi)
	\breve{f}(X_k|\mathscr{D}_1^k,\mathrm{E}_1^{k})
	\breve{f}(\Xi|\mathscr{D}_1^k,\mathrm{E}_1^{k}).
	\notag
\end{align}
Then, among these admissible pdfs, we seek the one that minimizes the KLD from $\breve{f}(B)$ to the FPD-ideal model
\begin{align}
	\label{eq:fI}
	&
	f_\mathrm{I}(B)
	\equiv
	f_\mathrm{I}
	(\mathrm{Y}_{k-w+1}^k,X_k,\Xi|\mathscr{D}_1^k,\mathrm{E}_1^{k-w}) 
	\\
	&
	\quad
	\equiv
	\!\!\!\!\!
	\prod_{q = k-w+1}^k 
	\!\!\!\!\!
	f(y_q|x_q,\Xi)
	f_\mathrm{I}(X_k|\mathscr{D}_1^k,\mathrm{E}_1^{k-w})
	f_\mathrm{I}(\Xi|\mathscr{D}_1^{k-1},\mathrm{E}_1^{k-w}).
	\notag
\end{align}
First, we outline how the transfer learning priors $f_\mathrm{I}(X_k | \mathscr{D}_1^k, \mathrm{E}_1^{k-w})$ and $f_\mathrm{I}(\Xi | \mathscr{D}_1^{k-1}, \mathrm{E}_1^{k-w})$ arise and are recursively updated. Afterwards, we proceed to solve the optimization problem \eqref{eq:FPD}. Note that the enforced conditional independence between the augmented state vector $X_k$ and the scale matrix $\Xi$ in \eqref{eq:varset} and \eqref{eq:fI} ensures the tractability and fixed computational complexity of $\hat{f}(B)$ and $f_\mathrm{I}(B)$.\pagebreak

The experimenter's initial knowledge about $X_1$ and $\Xi$ is provided at $k = 1$ in the form of the statistics $\big\{\hat{X}_{1|_0^0}, P_{1|_0^0}, \Sigma_0, \nu_0\big\}$ of the conjugate prior
$
f(X_{k}|\mathscr{D}_1^{k-1},\mathrm{E}_1^{k-w}) f_\mathrm{I}(\Xi|\mathscr{D}_1^{k-1},\mathrm{E}_1^{k-w}) \equiv \mathcal{N}\big(X_1 \big| \hat{X}_{1|_0^0}, P_{1|_0^0}\big)$ $\times i\mathcal{W}(\Xi | \Sigma_0, \nu_0 + w)
$ (§2.2.3.1 in \cite{Smi:05}).
The rationale for artificially increasing the degrees of freedom in the inverse-Wishart prior by $w$ is provided in the subsequent derivation (see the paragraph following \eqref{eq:fi_Xi_alg}).
To reflect the assumed structure of $\Xi$, $\Sigma_0$ is required to be diagonal.
Initialized with this prior, the recursion proceeds by repeatedly performing the target data step \eqref{eq:fi_DS}, the transfer learning step \eqref{eq:fi_TLS}--\eqref{eq:fi_Xistep}, and the time step \eqref{eq:fi_TS}. The resulting steps are given by:
\begin{align}
	\label{eq:fi_DS}
	&f_\mathrm{I}(X_k|\mathscr{D}_1^k,\mathrm{E}_1^{k-w})
	=
	\mathcal{N}\big(X_k\big|\hat{X}_{k|^k_{k-w}},P_{k|^k_{k-w}}\big)
	\\
	&
	\quad
	=
	f(y_{\mathrm{T};k}|x_k)
	f(X_k|\mathscr{D}_1^{k-1},\mathrm{E}_1^{k-w})
	/f(y_{\mathrm{T};k}|\mathscr{D}_1^{k-1},\mathrm{E}_1^{k-w}),
	\notag
	\\
	\label{eq:fi_TLS}
	&f(X_{k}|\mathscr{D}_1^{k},\mathrm{E}_1^{k-l})
	=
	\mathcal{N}\big(X_k\big|\hat{X}_{k|^k_{k-l}},P_{k|^k_{k-l}}\big)
	\\
	&
	\notag
	\quad
	\propto
	\exp\left\{
	\mathcal{E}_{\hat{f}(\Xi|\mathscr{D}_1^{k},\mathrm{E}_{1}^{k}) f_\mathrm{E}(y_{k-l}|y_{\mathrm{E};k-l},\Xi) }
	\bigl[\ln
	f(y_{k-l}|x_{k-l},\Xi)
	\bigr]
	\right\}
	\\
	&
	\quad
	\times f_\mathrm{I}(X_{k}|\mathscr{D}_1^{k},\mathrm{E}_1^{k-w}),
	\notag
	\\
	\label{eq:fi_Xistep}
	&f_\mathrm{I}(\Xi|\mathscr{D}_1^{k},\mathrm{E}_1^{k-l})
	=
	i\mathcal{W}(\Xi|\Sigma_{k-l},\nu_{k-l}+w)
	\\
	\notag
	&
	\quad
	\propto
	\exp\left\{
	\mathcal{E}_{\hat{f}(X_{k}|\mathscr{D}_1^{k},\mathrm{E}_{1}^{k}) f_\mathrm{E}(y_{k-l}|y_{\mathrm{E};k-l},\Xi) }
	\bigl[\ln
	f(y_{k-l}|x_{k-l},\Xi)
	\bigr]
	\right\}
	\\
	&
	\quad
	\times
	f_\mathrm{I}(\Xi|\mathscr{D}_1^{k-1},\mathrm{E}_1^{k-w}),
	\notag			
	\\
	\label{eq:fi_TS}
	&f(X_{k+1}|\mathscr{D}_1^{k},\mathrm{E}_1^{k-l}) =	\mathcal{N}\big(X_{k+1}\big|\hat{X}_{{k+1}|^k_{k-l}},P_{{k+1}|^k_{k-l}}\big)
	\\
	&
	\quad
	= 
	\left\{
	\!\!\!
	\begin{array}{ll}
		\int_{x^*}  
		f(x_{k+1}|x_{k},u_{k})
		f(X_{k}|\mathscr{D}_1^{k},\mathrm{E}_1^{k-l})
		\mathrm{d}x_{k-L}
		& \quad \!\! k > L \\[0.5em]
		f(x_{k+1}|x_{k},u_{k})
		f(X_{k}|\mathscr{D}_1^{k},\mathrm{E}_1^{k-l})
		& \quad \!\! k \leq L.
	\end{array}
	\right.
	\notag
\end{align}
Note that for $k \leq L$ (i.e., when $k - l = 0$) the external observation $y_{\mathrm{E};k-l}$ and the state $x_{k-l}$ appearing in \eqref{eq:fi_TLS}--\eqref{eq:fi_Xistep} are not available.
We consider incomplete information to be non-informative: 
$f_{\mathrm{E}}(y_{k-l}|y_{\mathrm{E};k-l},\Xi)\propto 1$, $f(y_{k-l}|x_{k-l},\Xi) \propto 1$. 
Consequently, the transfer learning step \eqref{eq:fi_TLS}--\eqref{eq:fi_Xistep} is effectively carried out only for $k > L$.

To perform the update \eqref{eq:fi_TLS}--\eqref{eq:fi_Xistep}, the optimization \eqref{eq:FPD} has to be solved. 
Given the ideal model \eqref{eq:fI}, the FPD-optimal choice of $\breve{f}(X_k|\mathscr{D}_1^k,\mathrm{E}_1^{k})$ and $\breve{f}(\Xi|\mathscr{D}_1^k,\mathrm{E}_1^{k})$ entering \eqref{eq:varset} is 
\begin{align}
	\label{eq:fhatX}
	&\hat{f}(X_k|\mathscr{D}_1^k,\mathrm{E}_1^{k})
	\propto
	\mathcal{N}\big(X_k\big|\hat{X}_{k|^k_{k-w}},P_{k|^k_{k-w}}\big)
	\\
	&
	\quad
	\times 
	\!\!\!\!\!
	\prod_{q=k-w+1}^k
	\!\!\!\!\!
	\exp{\left\{
		\mathcal{E}_{
			\hat{f}(\Xi|\mathscr{D}_1^k,\mathrm{E}_1^{k})	
		}
		\bigl[
		\ln(
		\mathcal{N}(y_{\mathrm{E};q}|Cx_q,R\Xi)
		)
		\bigr]
		\right\}},
	\notag
	\\
	\label{eq:fhatXi}
	&\hat{f}(\Xi|\mathscr{D}_1^k,\mathrm{E}_1^{k})
	\propto
	i\mathcal{W}(\Xi|\Sigma_{k-w},\nu_{k-w}+w)
	\\
	&
	\quad
	\times 
	\!\!\!\!\!
	\prod_{q=k-w+1}^k
	\!\!\!
	\frac{
		\exp{\left\{
			\mathcal{E}_{
				\hat{f}(X_k|\mathscr{D}_1^k,\mathrm{E}_1^{k})	
			}
			\bigl[
			\ln(
			\mathcal{N}(y_{\mathrm{E};q}|Cx_q,R\Xi)
			)
			\bigr]
			\right\}}
	}
	{
		\exp
		\Bigl\{
		\mathcal{E}_{
			\mathcal{N}(y_q|y_{\mathrm{E};q},\Xi)	
		}
		\bigl[
		\ln \left(
		\mathcal{N}(y_q|y_{\mathrm{E};q},\Xi)
		\right)
		\bigr]	
		\Bigr\}
	}
	.
	\notag
\end{align}
Here, the normal pdf $\mathcal{N}(y_{\mathrm{E};q}|Cx_{q},R\Xi)$ emerges from
\begin{align*}
	\mathcal{N}(y_{\mathrm{E};{{q}}}|Cx_{{q}},R\Xi)
	\propto
	\exp\Bigl\{
	\mathcal{E}_{f_\mathrm{E}(y_{{q}}|y_{\mathrm{E};{{q}}},\Xi)}
	\bigl[
		\ln(f(y_{{q}}|x_{{q}},\Xi))
	\bigr] 
	\Bigr\},
\end{align*}
and the product in the denominator of \eqref{eq:fhatXi} is reduced to
\begin{align*}
	\prod_{q=k-w+1}^k 
	\!\!\!\!\!
	\exp
	\Bigl\{
	\mathcal{E}_{
		\mathcal{N}(y_q|y_{\mathrm{E};q},\Xi)	
	}
	\bigl[		\ln \left(	\mathcal{N}(y_q|y_{\mathrm{E};q},\Xi)	\right)\bigr]
	\Bigr\}
	\propto
	\left| \Xi \right|^{-w \mathring{y}/2}.
\end{align*}
The result \eqref{eq:fhatX}--\eqref{eq:fhatXi} can be readily derived by decomposing the KLD \eqref{eq:KLD} for each variational marginal into parts dependent on and independent of that marginal.

The $\mathcal{N}(X_k)i\mathcal{W}(\Xi)$ form represents a self-consistent solution to the set of equations \eqref{eq:fhatX}--\eqref{eq:fhatXi}.
However, the mutually dependent shaping parameters \cite{Smi:05} of the normal and inverse-Wishart distributions are not given in a closed form.
Thus, we employ a tailored variant of the gradient descent iterative variational Bayesian (IVB) algorithm (Algorithm 1 in \cite{Smi:05}), which allows for conditional dependencies in $\breve{f}(B)$. 
Initializing \eqref{eq:est_Xi} and \eqref{eq:fi_Xi_alg} with $\Bigl\{   \hat{X}_{k|^k_k}^{[0]} \equiv  \hat{X}_{k|^k_{k-w}}, P_{k|^k_k}^{[0]}  \equiv  P_{k|^k_{k-w}}   {\Bigr\}}$, 
the \linebreak iterative updating procedure for $j = 1, \dots, N$ proceeds as $\;$
\linebreak
\vspace{-3.95mm}
\begin{align}
	\label{eq:est_X}
	&\mathcal{N}{\Bigl(}\!X_k\Big|\hat{X}_{k|^k_k}^{[j]},P_{k|^k_k}^{[j]}{\Bigr)}
	\propto
	\mathcal{N}{\Bigl(}\!X_k\Big|\hat{X}_{k|^k_{k-l}}^{[j]},P_{k|^k_{k-l}}^{[j]}{\Bigr)}
	\\[-0.3em]
	&
	\quad
	\times 
	\!\!\!\!\!
	\prod_{q=k-l+1}^k
	\!\!\!\!\!
	\exp{\left\{
		\mathcal{E}_{
			i\mathcal{W}(\Xi|\Sigma_{k}^{[j]},\nu_{k})	
		}
		\bigl[
		\ln(
		\mathcal{N}(y_{\mathrm{E};q}|Cx_q,R\Xi)
		)
		\bigr]
		\right\}},
	\notag
	\\
	\label{eq:est_Xi}
	&i\mathcal{W}{\bigl(}\Xi\big|\Sigma_{k}^{[j]},\nu_{k}{\bigr)}
	\propto
	i\mathcal{W}{\bigl(}\Xi\big|\Sigma_{k-l}^{[j]},\nu_{k-l}+w{\bigr)}
	\left| \Xi \right|^{w \mathring{y}/2}
	\\
	&
	\quad
	\times 
	\!\!\!\!\!
	\prod_{q=k-l+1}^k
	\!\!\!\!\!
	\exp{\left\{\!
		\mathcal{E}_{
			\mathcal{N}(X_k|\hat{X}_{k|^k_k}^{[j-1]},P_{k|^k_k}^{[j-1]})	
		}
		\bigl[
		\ln(
		\mathcal{N}(y_{\mathrm{E};q}|Cx_q,R\Xi)
		)
		\bigr]
		\!
		\right\}},
	\notag
\end{align}
where
\begin{align}
	\label{eq:fi_X_alg}
	&\mathcal{N}{\Bigl(} \!X_k\Big|\hat{X}_{k|^k_{k-l}}^{[j]},P_{k|^k_{k-l}}^{[j]}{\Big)}
	\propto
	\mathcal{N}{\bigl(}\!X_k\big|\hat{X}_{k|^k_{k-w}},P_{k|^k_{k-w}}{\bigr)}
	\\
	&
	\quad
	\times 
	\exp{\left\{
		\mathcal{E}_{
			i\mathcal{W}(\Xi|\Sigma_{k}^{[j]},\nu_{k})	
		}
		\bigl[
		\ln(
		\mathcal{N}(y_{\mathrm{E};k-l}|Cx_{k-l},R\Xi)
		)
		\bigr]
		\right\}},
	\notag
	\\
	\label{eq:fi_Xi_alg}
	&i\mathcal{W}{\bigl(}\Xi\big|\Sigma_{k-l}^{[j]},\nu_{k-l}+w{\bigr)}
	\propto
	i\mathcal{W}{\bigl(}\Xi\big|\Sigma_{k-w},\nu_{k-w}+w{\bigr)}
	\\
	&
	\quad
	\times 
	\exp{\left\{
		\!
		\mathcal{E}_{
			\mathcal{N}(X_k|\hat{X}_{k|^k_k}^{[j-1]},P_{k|^k_k}^{[j-1]})	
		}
		\bigl[
		\ln(
		\mathcal{N}(y_{\mathrm{E};k-l}|Cx_{k-l},R\Xi)
		)
		\bigr]
		\!
		\right\}}.
	\notag
\end{align}
Note that $i\mathcal{W}\bigl(\Xi\big|\Sigma_{k}^{[j]},\nu_{k}\bigr)$ retains the same degrees of freedom as $i\mathcal{W}\bigl(\Xi\big|\Sigma_{k-l}^{[j]},\nu_{k-w}+w\bigr)$, despite incorporating $w$ additional observations. This is caused by the term $\left| \Xi \right|^{w \mathring{y}/2}$ appearing in \eqref{eq:est_Xi}. 
The aim of artificially increasing the degrees of freedom in the FPD prior by $w$ is to compensate for this effect. As a result, the posterior statistics are updated consistently, enabling an unbiased transfer of information.

After the iterations, we assign 
$\Bigl\{
\hat{X}_{k|^k_k} 		\equiv \hat{X}_{k|^k_k}^{[N]}, 
\hat{X}_{k|^k_{k-l}} 	\equiv \hat{X}_{k|^k_{k-l}}^{[N]},
P_{k|^k_k} \equiv P_{k|^k_k}^{[N]},
P_{k|^k_{k-l}} \equiv P_{k|^k_{k-l}}^{[N]},
\Sigma_{k|^k_{k-l}} \equiv \Sigma_{k|^k_{k-l}}^{[N]} 
\Bigr\}$.
The FPD-optimal pdf $\mathcal{N}\bigl(X_k\big|\hat{X}_{k|^k_k},P_{k|^k_k}\bigr)$ is the result reported to the experimenter.
Apparently, the transfer learning step \eqref{eq:fi_TLS}--\eqref{eq:fi_Xistep} is resolved via \eqref{eq:fi_X_alg}--\eqref{eq:fi_Xi_alg} within the procedure for finding the FPD-optimal solution.
Note that only the resulting $\mathcal{N}\bigl(X_k\big|\hat{X}_{k|^k_{k-l}},P_{k|^k_{k-l}}\bigr)$ and $i\mathcal{W}\bigl(\Xi\big|\Sigma_{k-l},\nu_{k-l}\bigr)$ are propagated to the next step of the algorithm.
The decisions incorporating $y_{\mathrm{E};k-l}$ into these pdfs consider also the subsequent observations $\{\mathscr{D}_{k-l+1}^k, \mathrm{E}_{k-l+1}^k \}$, in addition to $\{\mathscr{D}_1^{k-l}, \mathrm{E}_1^{k-w} \}$ (see the expectations in \eqref{eq:fi_X_alg}--\eqref{eq:fi_Xi_alg}). Thus, the estimation error accumulation is mitigated to constitute a key advantage of FLISs in variational inference, as the filters typically rely solely on propagating the posterior $\hat{f}(X_k|\mathscr{D}_1^k,\mathrm{E}_1^{k})\hat{f}(\Xi|\mathscr{D}_1^k,\mathrm{E}_1^{k})$.

\subsection{Algebraic Recursion}
For brevity, let us introduce Lemma~\ref{lemma:SDU}, which generalizes the procedure for updating \( \mathcal{N}(X_k) \), given an observation.
To ensure unambiguous interpretation, we use generic notation in this formulation, as the lemma is referenced solely as a data-driven update procedure.
\begin{lemma}\label{lemma:SDU}
	Consider the observation \( z \in \mathbb{R}^{\mathring{z}} \) generated by the model \( \mathcal{N}(z|H x, \Gamma) \).
	Then, the initial belief \(\mathcal{N}(x|\mu_0,S_0)\) of the variable of interest \( x \) is updated using Bayes' rule:
	\begin{align}
		\label{eq:reviewer}
		\mathcal{N}(x|\mu_{\mathring{z}},S_{\mathring{z}}) \propto \mathcal{N}(x|\mu_0,S_0)\,\mathcal{N}(z|H x, \Gamma).
	\end{align}
	Assuming \(\Gamma \equiv \diag([\gamma_{\left[1\right]},\dots,\gamma_{\left[\mathring{z}\right]}]')\), the update based on the observation \( z \equiv [z_{\left[1\right]},\dots,z_{\left[\mathring{z}\right]}]' \) is preferably carried out by sequentially incorporating \(\mathcal{N}(z_{[i]}|H_{[i]}x,\gamma_{[i]})\) for \( i=1,\ldots,\mathring{z} \).
	Here, $H_{\left[i\right]} = \overline{\epsilon}_i^{\mathring{z}} H$ is the $i$th row of $H$. This matrix-inversion-free sequential data update (\(\mathrm{sdu}\)) procedure is summarized in Algorithm~\ref{alg:1}, which is referenced via
	\begin{align}\label{eq:sdu}
		\{\mu_{\mathring{z}}, S_{\mathring{z}}\} \equiv \mathrm{sdu}(\mu_{0},S_{0},H,\Gamma,z).
	\end{align}
	Throughout the article, we use the notation $\mathrm{sdu}(\cdot)$ interchangeably, with the specific inputs and outputs, in the defined order, naturally substituted based on the given context.
\begin{algorithm}[H]
	\caption{The sequential data update \eqref{eq:sdu} realizing the posterior update~\eqref{eq:reviewer}}
	\label{alg:1}
	\begin{algorithmic}[1]
		\State \textbf{Input:} $\mu_0,\,S_0,\,H,\,\Gamma,\,z$
		\For{$i \gets 1, \mathring{z}$}
		\State $H_{[i]} \gets \overline{\epsilon}_i^{\mathring{z}}H$
		\State $\gamma_{[i]} \gets \overline{\epsilon}_i^{\mathring{z}}\Gamma \epsilon_i^{\mathring{z}} $
		\State $K_{i} \gets S_{i-1}H_{[i]}'\big/\!\bigl(\gamma_{[i]}+H_{[i]}S_{i-1}H_{[i]}'\bigr)$
		\vspace{1pt}
		\State $\mu_{i} \gets \mu_{i-1}+K_{i}\bigl(\overline{\epsilon}_i^{\mathring{z}}z - H_{[i]}\mu_{i-1}\bigr)$
		\vspace{1pt}
		\State $S_{i} \gets \big(I_{\mathring{\mu}}\!\!-K_{i}H_{[i]}\big)S_{i-1}\big(I_{\mathring{\mu}}\!\!-K_{i}H_{[i]}\big)' + K_{i}\gamma_{[i]}K_{i}'$
		\EndFor
		\State \textbf{Output:} $\mu_{\mathring{z}},\,S_{\mathring{z}}$
	\end{algorithmic}
\end{algorithm}
\end{lemma}

By introducing the augmented output \eqref{eq:C_extended}, state transition \eqref{eq:A_extended}, and input and state noise \eqref{eq:B_extended} matrices
\begin{align}
	\label{eq:C_extended}
	\mathcal{C}_{k;q} &\equiv \overline{\epsilon}_{k-q+1}^w \otimes C, \quad q \in \{k-w+1,\dots,k\},
	\\
	\label{eq:A_extended}
	\mathcal{A}_{k+1} & \equiv 
	\Bigl[
	{\epsilon}_1^w \otimes A', 
	\begin{bmatrix}I_{l} & O_{l,w-l} \end{bmatrix}' \otimes I_{\mathring{x}}
	\Bigr]',
	\\
	\label{eq:B_extended}
	\mathcal{B}_{k+1}
	& \equiv
	\epsilon_1^{l+1} \otimes B,
	\qquad \;
	\!
	\mathcal{Q}_{k+1}	
	\equiv
	\diag(\epsilon_1^{l+1}) \otimes Q,	
\end{align}
we obtain the algebraic recursion via evaluating the ordered set of equations:
\begin{align}
	\label{eq:XP_DS}
	&
	\!\!\!
	\left\{
	\!
	\hat{X}_{k|_{k-w}^k}\!, P_{k|_{k-w}^k} 
	\!
	\right\}			
	\!=
	\mathrm{sdu}
	\Bigl(\!
	{\hat{X}}_{k|_{k-w}^{k-1}},
	P_{k|_{k-w}^{k-1}},
	\mathcal{C}_{k;k},
	R,
	y_{\mathrm{T};k}\!
	\Bigr),\!
	\\
	\label{eq:Sig:upd}
	&\Sigma_q^{[j]}
	=
	\Sigma_{q-1}^{[j]}
	\\
	&
	\quad + R^{-1} \circ
		\left[
		\diag\!\big(y_{\mathrm{E};q}- \mathcal{C}_{k;q} \hat{X}_{k|^k_{k}}^{[j-1]} )\big)^{2}\! + \mathcal{C}_{k;q} P_{k|^k_{k}}^{[j-1]} \mathcal{C}_{k;q}'
		\right],
	\notag
	\\
	\label{eq:XiHat}
	&\overline{\Xi}_k^{[j]}  \equiv 
	\Sigma_k^{[j]} \big/ (\nu_{k-w}+w),
	\\
	\label{eq:XP:update}
	&
	\!\!\!
	\left\{
	\!
	\hat{X}_{k|^k_{q}}^{[j]},
	P_{k|^k_{q}}^{[j]}
	\!\!\;
	\right\}			
	\!=
	\mathrm{sdu}
	\Bigl(\!
	\hat{X}_{k|^k_{q-1}}^{[j]}\!,
	P_{k|^k_{q-1}}^{[j]}\!,
	\mathcal{C}_{k;q}
	,
	R\overline{\Xi}_k^{[j]},
	y_{\mathrm{E};q}
	\!\Bigr),
	\\
	\label{eq:X_timestep}
	&{\hat{X}}_{k+1|_{k-l}^{k}}
	=\mathcal{A}_{k+1}{\hat{X}}_{k|_{k-l}^{k}}+ \mathcal{B}_{k+1}u_{k},
	\\
	\label{eq:P_timestep}
	&P_{k+1|_{k-l}^{k}}
	=\mathcal{A}_{k+1}P_{k|_{k-l}^{k}}\mathcal{A}_{k+1}^\prime+\mathcal{Q}_{k+1},
	\\
	\label{eq:nu}
	&\nu_{k-L} = \nu_{k-L-1} + 1, \qquad \quad \quad \; \! k>L.
\end{align}
The IVB initializers, to which \eqref{eq:Sig:upd} and \eqref{eq:XP:update} refer, are defined by
$\Bigl\{\Sigma_{k-w}^{[j]} \equiv \Sigma_{k-w}$, $\hat{X}_{k|^k_{k}}^{[0]}  =  \hat{X}_{k|^k_{k-w}}$, $P_{k|^k_{k}}^{[0]}  =  P_{k|^k_{k-w}}$, $\hat{X}_{k|^k_{k-w}}^{[j]} \equiv \hat{X}_{k|^k_{k-w}}$, $P_{k|^k_{k-w}}^{[j]} \equiv P_{k|^k_{k-w}}	\Bigr\}$. 
Note that \eqref{eq:Sig:upd} and \eqref{eq:XP:update} are evaluated for $q= k-w+1,\dots,k$.
The resulting Kalman FLIS with FPD-optimal knowledge transfer is summarized in Algorithm \ref{alg:FLIS}.

	\begin{algorithm}[h]
	\caption{Knowledge transfer for a Kalman FLIS}
	\label{alg:FLIS}
	\begin{algorithmic}[1]
		\State \textbf{Initialization:}
		\Statex Define the system $A, B, C$ and noise covariance $Q, R$ matrices and initialize the prior statistics ${\hat{X}}_{1|_0^0},P_{1|_0^0}, \Sigma_0,\nu_0$. 
		\State \textbf{Online estimation:}
		\For{$k \gets 1, \mathring{k}$}
		\State $l \gets \min{\left(L,k\right)}$, $w \gets \min{\left(L+1,k\right)}$
		
		\State \textbf{Input:}
		$ {\hat{X}}_{k|_{k-w}^{k-1}},$ $P_{k|_{k-w}^{k-1}},$ $ \Sigma_{k-w},\nu_{k-w},$ $ 
		u_{k},$ $ y_{\mathrm{T};k},$ $ y_{\mathrm{E};k}
		$
		
		\State Update: ${\hat{X}}_{k|_{k-w}^{k-1}} \rightarrow {\hat{X}}_{k|_{k-w}^{k}}, 
		P_{k|_{k-w}^{k-1}}	\rightarrow P_{k|_{k-w}^{k}}$ 
		\Comment{\raisebox{-1.25ex}{\shortstack[c]{\eqref{eq:C_extended} \\ \eqref{eq:XP_DS}}}}
		\State Initialize IVB: 
		$
		\hat{X}_{k|^k_{k}}^{[0]} \gets \hat{X}_{k|^k_{k-w}}, P_{k|^k_{k}}^{[0]} \gets P_{k|^k_{k-w}}
		$
		\For{$j\gets 1, N$}
		\State Initialize the sequential updating: 		
		$\Sigma_{k-w}^{[j]} \gets \Sigma_{k-w}$
		\Statex	$\hspace{1.05cm}  \hat{X}_{k|^k_{k-w}}^{[j]} \!\!\!\! \gets \hat{X}_{k|^k_{k-w}}, P_{k|^k_{k-w}}^{[j]} \!\!\!\! \gets P_{k|^k_{k-w}}$
		\For{$q\gets k-w+1,k$}
		\State Update: $\Sigma_{q-1}^{[j]} \rightarrow \Sigma_{q}^{[j]}$
		\Comment{\eqref{eq:C_extended}, \eqref{eq:Sig:upd}}
		\EndFor
		\State Assemble the matrix $\overline{\Xi}_k^{[j]}$ \Comment{\eqref{eq:XiHat}}
		\For{$q\gets k-w+1,k$}
		\State Update: 
		$\hat{X}_{k|^k_{q-1}}^{[j]} \!\! \rightarrow \hat{X}_{k|^k_{q}}^{[j]},
		P_{k|^k_{q-1}}^{[j]} \!\! \rightarrow P_{k|^k_{q}}^{[j]}$
		\Comment{\raisebox{-1.25ex}{\shortstack[c]{\eqref{eq:C_extended} \\\eqref{eq:XP:update}}}}
		\EndFor
		\EndFor	
		\State Update: 
		${\hat{X}}_{k|_{k-l}^{k}}  \rightarrow {\hat{X}}_{k+1|_{k-l}^{k}}, 
		$
		\Statex 
		$\hspace{0.6cm}   P_{k|_{k-l}^{k}}	 \rightarrow P_{k+1|_{k-l}^{k}}$
		\hskip\algorithmicindent\Comment{\eqref{eq:A_extended}, \eqref{eq:B_extended}, \eqref{eq:X_timestep}, \eqref{eq:P_timestep}}
		
		\State \textbf{Output:} 
		$
		\!
		\left\{
		\!
		\begin{array}{l} {\hat{X}}_{k|_k^k}, P_{k|_k^k}, {\hat{X}}_{k+1|_{k-l}^k}, P_{k+1|_{k-l}^k}, 
			\\
			\bigl(\bigl\{ \nu_{k-L},  \Sigma_{k-L} \bigr\}
			\;\; \textbf{if} \;\; k > L \bigr) 
		\end{array}
		\right.
		\!\!\!\!\!\!\!\!\!\!
		$
		\Comment{\eqref{eq:nu}}
		\EndFor
	\end{algorithmic}      
\end{algorithm}

	\section{Experiments}
This section provides a numerical example to empirically demonstrate the algorithm's performance.
The experiments consider the state-space model \eqref{eq:model_functional_y}--\eqref{eq:model_functional_x}, where the external data $y_{\mathrm{E};k}$ are generated by
\begin{align}\label{eq:ye_exp}
	y_{\mathrm{E};k} \sim \mathcal{N}(y_{\mathrm{E};k}|Cx_k,r_\mathrm{E} I_{\mathring{y}}),
\end{align}
similarly to \cite{Pap:19}. 
The variable coefficient $r_\mathrm{E}$ is a simulation parameter that defines how precise (or imprecise) the external knowledge is. Thus, by varying its value, we evaluate the robustness and sensitivity
in different scenarios.
The qualities of the designed algorithm are assessed in two regimes that highlight the merits of the knowledge transfer FLIS (TFLIS) concept.
The first regime reports \eqref{eq:smot} as the estimation result, and we refer to this one as the \textit{smoothing regime} (TFLIS-S). 
The result of the other one is given by \eqref{eq:filt}, and we refer to it as the \textit{filtering regime} (TFLIS-F). We have
\begin{align}
	\label{eq:smot}
	\hat{x}_{k|_{k+L}^{k+L}} &\equiv \bigl(\overline{\epsilon}_{L+1}^{L+1} \otimes I_{\mathring{x}}\bigr)\,\hat{X}_{k+L|_{k+L}^{k+L}},
	\\
	\label{eq:filt}
	\hat{x}_{k|_k^k} &\equiv \bigl(\overline{\epsilon}_1^{w} \otimes I_{\mathring{x}}\bigr)\,\hat{X}_{k|_k^k}.
\end{align}
A comparison is made with the Kalman filter (KF) \cite{Kal:60} and the fixed-lag smoother (FLS) \cite{Moore:73}, which, given the stochastic model \eqref{eq:ye_exp} explicitly, treat \( y_{\mathrm{E};k} \) in the same way as \( y_{\mathrm{T};k} \) (i.e., by directly incorporating external knowledge via Bayes' rule).	
The comparison includes the \textit{``Static Bayesian transfer learning filter with scale relaxation''} (RSTF) (Algorithm 1 in \cite{Pap:19}) and an isolated (processing no external data) Kalman filter (iKF) and fixed-lag smoother (iFLS).

The state estimation accuracy is evaluated by calculating the squared error $\mathrm{SE}_k$ and the mean squared error $\mathrm{MSE}$ between the true state and its estimate: 	
\begin{align*}
	\mathrm{SE}_k &= \norm{x_k - \hat{x}_{k|^t_e}}_2^2, \quad
	& \mathrm{MSE} &= \frac{1}{\mathring{k}-L} \sum_{k=1}^{\mathring{k}-L} \mathrm{SE}_k.
\end{align*}
The indices $t$ and $e$ for the considered methods are given by Tab. \ref{tab:1}, and the simulation length is set to $\mathring{k}=50$. 
\setlength{\jot}{0pt}
\begin{table}[H]	
	\centering
	\caption{The estimates $\hat{x}_{k|^t_e}$ of $x_k$ provided by the compared algorithms}
	\label{tab:1}
	\begin{tabularx}{\columnwidth}{>{\centering\arraybackslash}X
			>{\centering\arraybackslash}X
			>{\centering\arraybackslash}X
			>{\centering\arraybackslash}X
			>{\centering\arraybackslash}X}
		\Xhline{1pt}
		\makecell[c]{iFLS} & 
		\makecell[c]{iKF} & 
		\makecell[c]{FLS,\\ TFLIS-S} & 
		\makecell[c]{KF,\\ TFLIS-F} & 
		\makecell[c]{RSTF} \\
		\hline
		$\begin{matrix} \begin{aligned} t &= k+L \\ e &= 0 \end{aligned} \end{matrix}$ &
		$\begin{matrix} \begin{aligned} t &= k \\ e &= 0 \end{aligned} \end{matrix}$ & 
		$\begin{matrix} \begin{aligned} t &= k+L \\ e &= k+L \end{aligned} \end{matrix}$ &
		$\begin{matrix} \begin{aligned} t &= k \\ e &= k \end{aligned} \end{matrix}$ &
		$\begin{matrix} \begin{aligned} t &= k \\ e &= k - 1 \end{aligned} \end{matrix}$ \\
		\Xhline{1pt}
	\end{tabularx}
\end{table}

We consider a discretized position-velocity system (§7.3.1 in \cite{Sim:06}), stimulated by acceleration as the input and extended to include the observations of both the position and the velocity. The system and noise covariance matrices are:
\begin{align*}
	\begin{split}
		A &= \begin{bmatrix}
			1 &  1 \\[1pt] 0 & 1
		\end{bmatrix},
		\quad
		B = \begin{bmatrix}
			0.5 \\[1pt] 1
		\end{bmatrix},
		\quad
		Q = 10^{-4} \begin{bmatrix}
			0.25 & 0.5\\[1pt] 0.5 & 1
		\end{bmatrix},
		\\[0.4ex]
		C &= I_2,
		\hspace{33.5pt}
		R = 10^{-3}\,I_2.
	\end{split}
\end{align*}
The true initial state $x_1 \in \left[-0.05, 0.05\right]^2$ is generated randomly, following a uniform pdf. The prior statistics are set to
${\Bigl\{}\hat{X}_{1|_0^0}={O_{2,1}},  P_{1|_0^0} = 10^{7}I_{2}, \nu_0 = 0, \Sigma_0 = O_{2,2}{\Bigr\}}.$
Similarly, for the RSTF we choose the ``\textit{R3}'' setting discussed in \cite{Pap:19}, obtaining a transfer influenced only by the data.
The input sequence \(\{u_k\} \in \{-1, 1\}\)
is generated by a maximum-length pseudorandom binary sequence generator based on a 4-bit shift register (§3.2 in \cite{Gol:67}), with the seed randomly chosen from 15 valid variants under a uniform distribution.
Accordingly, the original RSTF algorithm \cite{Pap:19} had to be refined to account for the presence of an input, $u_k$. As mentioned earlier, the RSTF considers a pair of Kalman filters, one for the target system and the other for the external one. To ensure consistent conditions with respect to the TFLIS and the setup to generate external data \eqref{eq:ye_exp}, we assume that both systems are stimulated by the same input, $u_k$. However, note that the homogeneous excitation of the systems is, in general, not strictly required.
The fixed lag is set to $L=2$, and the number of IVB iterations is $N=10$ for both knowledge transfer algorithms.

Figure \ref{fig:01}(a) illustrates the algorithms' performance related to the precision of the externally supplied data, determined by the coefficient $r_\mathrm{E}$, and Fig. \ref{fig:01}(b) shows the time evolution of $\mathrm{SE}_k$ for $r_\mathrm{E} = 10^{-3}$.
The $\mathrm{SE}_k$ and MSE levels of the iKF and iFLS serve as 
a reference to determine whether the transfer learning is positive or negative. The KF and FLS provide, owing to the exact knowledge of the stochastic relation between \( y_{\mathrm{E};k} \) and \( x_k \), the optimal solution to the estimation problem.
Figure~\ref{fig:01}(a) shows that the TFLIS-F maintains a robustness similar to the RSTF when the external data are imprecise but, advantageously, offers a substantial benefit under precise external observations.
Moreover, Fig.~\ref{fig:01}(b) illustrates that the TFLIS-F and TFLIS-S gradually approach the optimal solution, in contrast to the RSTF.
Expectably, the smoothing algorithms iFLS, FLS, and TFLIS-S provide a higher accuracy compared to their filtering counterparts, the iKF, KF, and TFLIS-F, respectively, at the cost of delayed estimates by \( L \).

Remarkably, the TFLIS-F might exhibit a slight negative transfer at the early time steps. With imprecise external knowledge, the algorithm may initially fail, increasing $\overline{\Xi}_k^{\left[N\right]}$ sufficiently to
reject inaccurate external observations due to the limited amount of information collected. However, thanks to the FLIS structure, the accumulation of this initial error is significantly suppressed. In the \textit{smoothing regime},  this issue
does not arise, as all reported estimates of the TFLIS-S are already refined using future observations.

\definecolor{MyLightBlue}{HTML}{11A0DB}
\definecolor{MyMidBlue}{HTML}{004393}
\definecolor{MyGreen}{HTML}{67BC4C}	
\definecolor{MyPurp}{HTML}{CC1398}
\definecolor{MyRed}{HTML}{ff776e}

\begin{figure}[!t]
	\begin{subfigure}{\linewidth}
		\begin{tikzpicture}
			\begin{axis}[
				tick scale binop=\times,
				ymode=normal, xmode=log, height=6.6cm, width=\linewidth+0.5cm,
				xlabel=$r_\mathrm{E}$, ylabel=$\mathrm{MSE}$,
				xmin=1E-6, ymin=0, xmax=1E0, ymax=0.00075,
				x label style={at={(axis description cs:0.5,-0.028mm)}, anchor=north},
				ylabel={$\mathrm{MSE}$},
				y label style={at={(axis description cs:-0.020mm,0.5)}, anchor=south},
				title=(a),
				title style={at={(0.5,1)}, anchor=center},
				xmajorgrids=true, ymajorgrids=true, major grid style={thin,color=gray,dotted},
				legend entries={iFLS, iKF, FLS, KF, TFLIS-S, TFLIS-F, RSTF-R3},
				legend style={at={(0.000,1)}, anchor=north west, cells={anchor=west},font=\footnotesize},
				ytick distance=0.0001,
				xminorgrids=true, minor grid style={thin,color=gray,dotted},
				minor tick num=2,
				]
				\addplot[solid, color=MyPurp, line width=1.5pt] table[x=Rs,y=FLKS_target] {MSNE_all.txt}; 
				\addplot[densely dashed, color=MyPurp, line width=1.5pt] table[x=Rs,y=KF_target] {MSNE_all.txt}; 
				\addplot[solid, color=Black, line width=1.5pt] table[x=Rs,y=FLKS_both] {MSNE_all.txt}; 
				\addplot[densely dashed, color=Black, line width=1.5pt] table[x=Rs,y=KF_both] {MSNE_all.txt}; 
				\addplot[solid, color=MyLightBlue, line width=1.5pt] table[x=Rs,y=My_at_k-L] {MSNE_all.txt}; 
				\addplot[densely dashed, color=MyLightBlue, line width=1.5pt] table[x=Rs,y=My_at_K] {MSNE_all.txt}; 
				\addplot[densely dashed, color=MyGreen, line width=1.5pt] table[x=Rs,y=Milan_RTFL] {MSNE_all.txt}; 
			\end{axis}
		\end{tikzpicture}
	\end{subfigure}
	\\
	\begin{subfigure}{\linewidth}
		\begin{tikzpicture}
			\begin{axis}[
				tick scale binop=\times,
				ymode=normal, xmode=normal, height=4cm, width=\linewidth+0.5cm,
				xlabel=$k$, ylabel=$\mathrm{MS}_k$,
				xmin=1, xmax=48, ymin=0, ymax=0.002,
				xlabel={$k$},
				x label style={at={(axis description cs:0.5,-0.04mm)}, anchor=north},
				ylabel={$\mathrm{MS}_k$},
				y label style={at={(axis description cs:-0.0175mm,0.5)}, anchor=south},
				xtick={1,10,20,30,40,48},
				title=(b),
				title style={at={(0.5,1)}, anchor=center},
				xmajorgrids=true, ymajorgrids=true, major grid style={thin,color=gray,dotted},
				]
				\addplot[solid, color=MyPurp, line width=1.5pt] table[x=k, y=FLKS_target] {Rs_1E-3.txt}; 
				\addplot[densely dashed, color=MyPurp, line width=1.5pt] table[x=k, y=KF_target] {Rs_1E-3.txt}; 
				\addplot[solid, color=Black, line width=1.5pt] table[x=k, y=FLKS_both] {Rs_1E-3.txt}; 
				\addplot[densely dashed, color=Black, line width=1.5pt] table[x=k, y=KF_both] {Rs_1E-3.txt}; 
				\addplot[solid, color=MyLightBlue, line width=1.5pt] table[x=k, y=My_at_k-L] {Rs_1E-3.txt}; 
				\addplot[densely dashed, color=MyLightBlue, line width=1.5pt] table[x=k, y=My_at_K] {Rs_1E-3.txt}; 
				\addplot[densely dashed, color=MyGreen, line width=1.5pt] table[x=k, y=Milan_RTFL] {Rs_1E-3.txt}; 
			\end{axis}
		\end{tikzpicture}
	\end{subfigure}
	\caption{
		(a) The state estimation MSE versus the external observation variance $r_\mathrm{E}I_2$;  
		(b) the time evolution of the state estimate $\mathrm{SE}_k$ for $r_\mathrm{E} =$  $10^{-3}$.
		The results are given as the average of 10000 independent simulation runs, each of a length $\mathring{k}=50$.			
		The compared algorithms are:
		(i) Isolated (no transfer) fixed-lag smoother (iFLS) \cite{Moore:73}; 
		(ii) Isolated Kalman filter (iKF) \cite{Kal:60};
		(iii) Fixed-lag smoother (FLS) \cite{Moore:73}, given the exact external observation model \eqref{eq:ye_exp};
		(iv) Kalman filter (KF) \cite{Kal:60}, given the exact external observation model \eqref{eq:ye_exp};
		(v) Knowledge transfer for a Kalman FLIS in the \textit{smoothing regime} (TFLIS-S), given by Algorithm \ref{alg:FLIS} and \eqref{eq:smot}, with fixed lag $L = 2$; 
		(vi) Knowledge transfer for a Kalman FLIS in the \textit{filtering regime} (TFLIS-F), given by Algorithm \ref{alg:FLIS} and \eqref{eq:filt}, with $L = 2$; 
		(vii) ``\textit{Static Bayesian transfer learning filter with scale relaxation}'' in Regime 3 (RSTF-R3) \cite{Pap:19}.
	}
	\label{fig:01}
\end{figure}
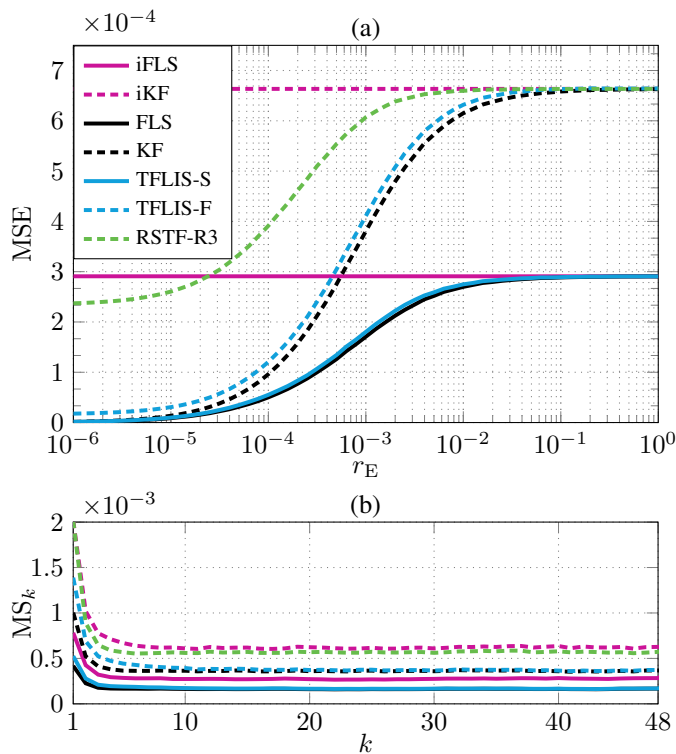

\section{Conclusion}
The article proposes an online FPD-optimal knowledge transfer mechanism for a Kalman FLIS. 
Compared to \cite{Pap:19}, external information is incorporated directly, avoiding further processing, and, importantly, obviating the need for any additional \pagebreak external knowledge besides the output measurements.
Owing to the transfer design and FLIS structure, the designed algorithm better exploited the precise external data while retaining its robustness in rejecting the imprecise ones.
The experimental results indicate that this research contributes to the field of probabilistic knowledge transfer in state estimation, with promising implications for real-world applications.

The future work will focus on applying our design to the estimation of the electric current in the stator coils of interior permanent magnet synchronous motors. Magnetic flux sensors, which embody an external source of knowledge, are desired to improve the estimation properties while obviating the necessity to know their relation to the currents explicitly.


\begin{thebibliography}{00}
	\bibitem{Zhu:21}
	F. Zhuang et al., ``A Comprehensive Survey on Transfer Learning,''  in \emph{Proceedings of the IEEE}, vol. 109, pp. 43--76, Jan. 2021.
	
	\bibitem{Torr:09}
	L. Torrey, J. Shavlik, ``Transfer learning,'' in \emph{Handbook of Research on Machine Learning, Applications and Trends: Algorithms, Methods and Techniques}, pp. 242--264, 2010.
	
	\bibitem{Zhu:23}
	Z. Zhu et al., ``Transfer Learning in Deep Reinforcement Learning: A Survey,'' in \emph{IEEE Transactions on Pattern Analysis and Machine Intelligence}, vol. 45, no. 11, pp. 13344--13362, 1 Nov. 2023.
	
	\bibitem{Karn:16}
	A. Quinn, M. Kárný, and T. V. Guy, ``Fully probabilistic design of hierarchical Bayesian models,'' in \emph{Information Sciences}, vol. 369, pp. 532--547, 2016.
	
	\bibitem{Karn:96}
	M. Kárný, ``Towards fully probabilistic control design,'' in \emph{Automatica}, vol. 32, no. 12, pp. 1719--1722, 1996.
	
	\bibitem{Guy:12}
	M. Karny, T.V. Guy, ``On Support of Imperfect Bayesian Participants,'' in: \emph{Decision Making with Imperfect Decision Makers}, pp. 29--56, 2012.
	
	
	\bibitem{Sho:80}
	J. Shore and R. Johnson, ``Axiomatic derivation of the principle of maximum entropy and the principle of minimum cross-entropy,'' in \emph{IEEE Transactions on Information Theory}, vol. 26, no. 1, pp. 26--37, 1980.
	
	\bibitem{Kull:51}
	S.~Kullback and R.~A.~Leibler, ``On information and sufficiency,'' in  \emph{The Annals of Mathematical Statistics}, vol. 22, no. 1, pp. 79--86, 1951.
	
	\bibitem{Quinn:18}
	C. Foley and A. Quinn, ``Fully probabilistic design for knowledge transfer in a pair of Kalman filters,'' in \emph{IEEE Signal Processing Letters}, vol. 25, no. 4, pp. 487--490, 2018.
	
	\bibitem{Pap:18}
	M. Papež and A. Quinn, ``Dynamic Bayesian knowledge transfer between a pair of Kalman filters,'' in \emph{2018 IEEE 28th International Workshop on Machine Learning for Signal Processing}.  Aalborg, Denmark.
	
	\bibitem{Pap:19}
	M. Papež and A. Quinn, ``Robust Bayesian transfer learning between Kalman filters,'' in \emph{2019 IEEE 29th International Workshop on Machine Learning for Signal Processing}, Pittsburgh, PA, USA.
	
	\bibitem{Pap:21}
	M. Papež and A. Quinn, ``Hierarchical Bayesian Transfer Learning Between a Pair of Kalman Filters,'' in \emph{2021 32nd Irish Signals and Systems Conference}, Athlone, Ireland.
	
	\bibitem{Pap:stud}
	M. Papež and A. Quinn, ``Bayesian transfer learning between Student-t filters,'' in \emph{Signal Processing}, vol.~175, 2020, Art. no.~107624.
	
	\bibitem{Kal:60}
	R. E. Kalman, ``A new approach to linear filtering and prediction problems,'' in \emph{Transactions of the ASME-Journal of Basic Engineering}, vol. 82, no. 1, pp. 35--45, 1960.
	
	\bibitem{RTS:65}
	H. E. Rauch, F. Tung, and C. T. Striebel, ``Maximum likelihood estimates of linear dynamic systems,'' in \emph{AIAA Journal}, vol. 3, no. 8, pp. 1445--1450, 1965.
	
	\bibitem{Moore:73}
	J. B. Moore, ``Discrete-time fixed-lag smoothing algorithms,'' in \emph{Automatica}, vol. 9, no. 2, pp. 163--173, 1973.
	
	\bibitem{Xu:21}
	H. Xu, K. Duan, H. Yuan, W. Xie and Y. Wang, ``Adaptive Fixed-Lag Smoothing Algorithms Based on the Variational Bayesian Method,'' in \emph{IEEE Trans. Automat. Control}, vol. 66, no. 10, pp. 4881--4887, 2021.
	
	\bibitem{Kol:33}
	A. Kolmogorow, ``Grundbegriffe der Wahrscheinlichkeitsrechnung''. Springer, Berlin 1933, Reprint 1974.
	
	
	
	\bibitem{Med:67}
	J. S. Meditch, ``On Optimal Linear Smoothing Theory,'' in \emph{Information and Control}, vol. 10, pp. 598--615, 1967.
	
	
	
	\bibitem{Karn:17}
	A. Quinn, M. Kárný and T. V. Guy, ``Optimal design of priors constrained by external predictors,'' in \emph{International Journal of Approximate Reasoning}, vol. 84, pp. 150--158, 2017.
	
	\bibitem{Smi:05}
	V. Šmídl and A. Quinn, ``The Variational Bayes Method in Signal Processing.'' Springer, 2005.
	
	\bibitem{Sim:06}
	D. Simon, ``Optimal State Estimation: Kalman, H Infinity, and Nonlinear Approaches.'' John Wiley \& Sons, 2006.
	
	\bibitem{Gol:67}
	S. W. Golomb, ``Shift Register Sequences.'' Holden-Day, San Francisco 1967.
	
\end{thebibliography}
\end{document}